\documentclass[useAMS,usenatbib]{mn2e}
\usepackage{txfonts,graphicx,natbib}
\bibpunct{(}{)}{;}{a}{}{,}

\begin{document}

\title[SN 2005cs in M51]{SN 2005cs in M51\\ I. The first month of evolution of a subluminous SN
II plateau} \author[Pastorello et al.]{A. Pastorello$^{1}$
\thanks{e--mail: pasto@MPA-Garching.MPG.DE}, 
D. Sauer$^{2}$, S. Taubenberger$^{1}$, P. A. Mazzali$^{1,2,3,4}$, K. Nomoto$^{3,4}$,\and
 K. S. Kawabata$^{5}$, S. Benetti$^{6}$, N. Elias--Rosa$^{6}$, A. Harutyunyan$^{6}$, H. Navasardyan$^{6}$,\and
 L. Zampieri$^{6}$, T. Iijima$^{7}$, M. T. Botticella$^{8,9}$, G. Di Rico$^{8,9}$, M. Del Principe$^{9}$,\and
 M. Dolci$^{9}$, S. Gagliardi$^{9}$, M. Ragni$^{8,9}$,
G. Valentini$^{9}$\\
$^{1}$ Max-Planck-Institut f\"{u}r Astrophysik,
Karl-Schwarzschild-Str. 1, 85741 Garching bei M\"{u}nchen, Germany\\
$^{2}$ INAF Osservatorio Astronomico di Trieste, Via Tiepolo 11, 34131 Trieste, Italy\\
$^{3}$ Department of Astronomy, School of Science, University of Tokyo, Bunkyo-ku, Tokyo 113-0033, Japan\\
$^{4}$ Research Center for the Early Universe, School of Science, University of Tokyo, Bunkyo-ku, Tokyo 113-0033, Japan\\
$^{5}$  Hiroshima Astrophysical Science Center, Hiroshima University, Hiroshima 739-8526, Japan\\
$^{6}$  INAF Osservatorio Astronomico di Padova, Vicolo dell'Osservatorio 5, 35122 Padova, Italy \\
$^{7}$  INAF Osservatorio Astronomico di Padova, Sezione di Asiago, Via dell'Osservatorio 8, 36012 Asiago (Vicenza), Italy \\
$^{8}$  Universit\`a degli Studi di Teramo, Viale Crucioli 122, 64100 Teramo, Italy\\
$^{9}$  INAF Osservatorio Astronomico di Collurania, via M. Maggini, 64100 Teramo, Italy}
%Dipartimento di Scienze della Comunicazione, Universitá di Teramo, viale Crucioli 122, 64100 Teramo, Italy

\date{Accepted
.....; Received ....; in original form ....}

\maketitle

\begin{abstract}
Early time optical observations of supernova (SN)
2005cs in the Whirlpool Galaxy (M51), are reported. 
Photometric data suggest that SN 2005cs is a moderately under--luminous Type II plateau
supernova (SN IIP). The SN was unusually blue at early
epochs (U--B $\approx-$0.9 about three days after
explosion) which indicates very high continuum temperatures. 
The spectra show relatively narrow P--Cygni
features, suggesting ejecta velocities lower
than observed in more typical SNe IIP. The earliest spectra show weak
absorption features in the blue wing of the He I 5876\AA~ absorption component
and, less clearly, of H$\beta$ and H$\alpha$. 
Based on spectral modelling, two different interpretations
can be proposed: these features may either be due to high--velocity H and He I
components, or (more likely) be produced by different ions (N II, Si II).
Analogies with the low--luminosity, $^{56}$Ni--poor, low--velocity SNe IIP
are also discussed.
 While a more extended spectral coverage is necessary in order to determine
accurately the properties of the progenitor star, 
published estimates of the progenitor mass seem not 
to be consistent with stellar evolution models.
\end{abstract}


\section{Introduction}  \label{intro}
Type II supernovae (SNe II) are believed to be produced by the explosion following the
core--collapse of massive stars that retained most of their H envelope
at the time of explosion. Some SNe II spend a period at almost constant 
luminosity: this phase, lasting sometimes a few
months, is called ``plateau'' (hence the label SN IIP).
When the SN enters this phase, the
temperature is low enough that the massive H envelope, initially
ionized because of the deposition of energy by the shock wave,
starts to recombine. 
After recombination, the light curve of SNe IIP declines steeply, 
until it settles onto the ``radioactive tail'', as do other SN Types. 
In this phase the luminosity is
due mainly to the radioactive decay of $^{56}$Ni to $^{56}$Co to
$^{56}$Fe.

Despite the considerable number of SNe IIP studied in recent
years \citep[e.g.][]{pata94,ham03}, the physical properties of the
progenitor stars are still a matter of debate.
Thanks to the direct identification of several SN precursors in deep
pre--explosion images,
important constraints have been established on the nature
of the progenitors of SNe IIP. The first SN with a known progenitor
was SN 1987A in the Large Magellanic Cloud.
Archival images showed that the supergiant progenitor was unusually blue
\citep[e.g.][]{sonn87}.
The present ensemble of SNe IIP with detected
progenitor includes SN 2003gd \citep{van03,smar04,maggie04},
SN 2004et \citep{li05a}, SN 1999ev \citep{mau05a} and, now, SN 2005cs
\citep{mau05b,li05b}.
Lower magnitude limits or ambiguous detections have
been obtained for other SNe IIP precursors \citep[e.g. SN 1999br,][]{mau05a}.
All these observations  seem to support the idea that most  SNe IIP originate
from the explosion of moderately massive stars (M $\leq$ 15$_{\odot}$).

\begin{figure}
 \resizebox{\hsize}{!}{\includegraphics{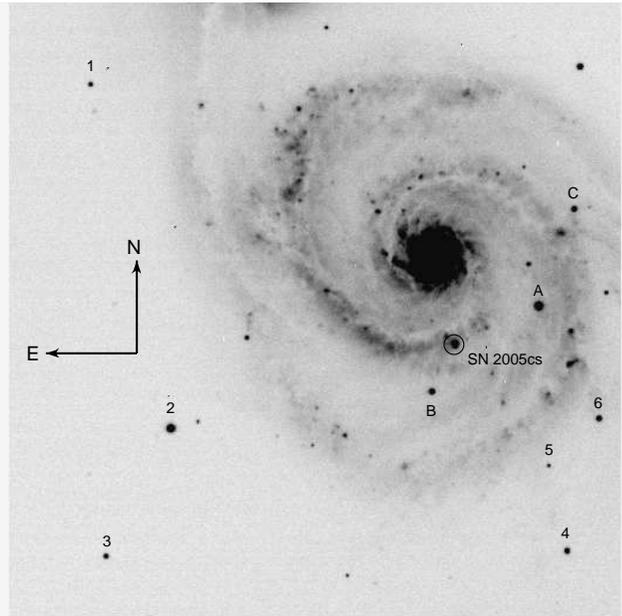}}
   \caption{SN 2005cs in M51: V band image obtained on July 1, 2005 with
the Copernico 1.82m telescope of Mt. Ekar, Asiago (Italy).
The sequence stars from \citet{ric96} are
labelled with letters. Some of our local sequence standards (Pastorello et
   al. in prep.) are
 indicated by numbers.}
   \label{fig:M51field}
\end{figure}

 SN 2005cs was discovered in the famous Whirlpool 
Galaxy (NGC 5194 or M51) by \citet{klo05} on 2005 June 28.905 UT. 
\citet{modj05} classified it as a young Type II SN.
The earliest detection was that of M. Fiedler on June 27.91 UT
(SNWeb\footnote{\it http://www.astrosurf.com/snweb2/}).
Nothing was visible on June 20.6 UT \citep{klo05} at the SN position.
Moreover, no clear detection of SN 2005cs was found on several
different images obtained on June 26 by other amateur observers (SNWeb). 
In particular, the SN site was
monitored by the team of the Osservatorio Astronomico
``Geminiano Montanari'' (Cavezzo, Modena, Italy) on June 26.89
using a Newton 0.4m telescope and nothing was detected below
the following limits: B $\geq$ 17.3, V $\geq$ 17.7, R $\geq$ 17.6.
These detection limits constrain the explosion time
to a very small uncertainty (about 1 day). Therefore 
in this paper we adopt
June 27.5 UT (JD = 2453549$\pm$1) as the explosion epoch .

The coordinates of SN 2005cs are $\alpha = 13^{h}29^{m}52\fs85$ and $\delta  = +47\degr10\arcmin36\farcs3$
(J2000). The object lies in the southern arm of M51, 15" West and
67".3 South of the galaxy
nucleus (Fig. \ref{fig:M51field}). M51 is classified by NED\footnote{\it
http://nedwww.ipac.caltech.edu/} as an SA(s)bc peculiar galaxy. 
The galaxy also hosted the well studied core--collapse
SN 1994I \citep{whee94,fili95,ric96,cloc96}.
% and the historical Type I
%SN 1945A exploded in the companion galaxy NGC 5195 \citep{vand99}.

A candidate progenitor for SN 2005cs was identified in combined HST ACS F814W
images ($\sim$ I band)  as a red supergiant
(M$_{ZAMS}$ = 9$^{+3}_{-2}$ M$_{\odot}$) of spectral type in the range
K0--M4 \citep{li05b,mau05b}. 
These two papers report different values for the I band detection magnitude:
24.15 (i.e. absolute magnitude M$_I \approx -$5.5) and 23.3 (M$_I \approx -$6.4, adopting the same
reddening and distance as \citet{li05b}), respectively.
The position of the candidate as measured by \citet{li05b} is 
$\alpha = 13^{h}29^{m}52\fs76$, $\delta = +47\degr10\arcmin36\farcs11$ (J2000).
Alternatively, \citet{ric05} claims that SN~2005cs exploded near a
cluster of young stars and finds that the progenitor candidate could be
a blue star at $\alpha = 13^{h}29^{m}52\fs803$, $\delta =
+47\degr10\arcmin36\farcs52$ (J2000), with M$_V$ $\approx -$6.
However, since SN~2005cs is evolving like a normal plateau event (see Sect. \ref{sect:photo}),
this blue supergiant progenitor candidate is not convincing.
%If we believe in the progenitor identification performed by
%\citet{li05b} and \citet{mau05b}, this could provide important 
%clues in favour of a lower--mass scenario \citep[e.g.][]{chug00}
%for progenitors of most Type IIP SNe. 
%including the under--energetic,
%$^{56}$Ni--poor events like SN~1997D \citep[e.g.][]{pasto04}.
%However, as we will show in Sect. \ref{discussion}.

Here we present early time optical observations of SN 2005cs
(until $\sim$ 1 month after the explosion). In Sect. 2 we describe the
photometric evolution of SN 2005cs and in Sect. 3 we analyse the
spectroscopic data. A discussion follows in Sect. 4. 

\section{Photometry} \label{sect:photo}

Our photometric data were obtained using seven different
telescopes, and cover 19 epochs (including the prediscovery limit),
until approximately 35 days after the explosion. \\
All data were pre--reduced with standard IRAF\footnote{IRAF is
distributed by the National Optical Astronomy Observatories, which are 
operated by the Association of Universities for Research in Astronomy,
Inc, under contract to the National Science Foundation.} procedures,
and instrumental SN magnitudes were determined using the
point--spread function (PSF) fitting technique performed with the
``SNOOPY''\footnote{SNOOPY is a package implemented in
IRAF by E. Cappellaro, based on DAOPHOT.} package. Since SN 2005cs
is a bright object, this technique provides acceptable results,
although the background region of SN 2005cs is extremely complicated, 
such that the subtraction of the template could be more appropriate
when the SN fades.

In order to transform instrumental magnitudes to the standard
Johnson--Cousins system, first--order colour corrections were
applied with colour terms derived from observations of photometric
standard fields \citep{land92}. 
The photometric zeropoints were finally determined by comparing the
magnitudes of a local sequence of stars in the vicinity of M51 (cf. Fig. \ref{fig:M51field}) 
to the values reported for some of these stars by \citet{ric96} in
their study on SN 1994I. 
%These measurements provide sufficient arguments for the
%discussion in this paper, but 
A complete definitive photometry
(calibrated on a larger sequence of stars, i.e. those labelled by
numbers in Fig. \ref{fig:M51field}) 
will be presented in a forthcoming paper \citep{pasto05}.

\begin{table*}
%\begin{footnotesize}
\caption{UBVRI magnitudes of SN 2005cs and assigned errors.
Both measurement errors and uncertainties in the photometric
calibration (the most important source of errors) are taken into account. 
Measurement uncertainties give a minor contribution to the total error
because, despite the relatively complex background, the SN was much brighter than
any other nearby source.
}
\centering
\label{SN_mags}
\begin{tabular}{cccccccc}
\hline\hline
Date & JD & $U$ & $B$ & $V$ & $R$ & $I$ & Inst.\\
& (+2400000) & & & & & &\\ \hline
05/06/26 & 53548.39 &                & $\geq$ 17.3      & $\geq$ 17.7      & $\geq$ 17.6 &  &NCO\\ 
05/06/30 & 53552.36 & 13.48$\pm$0.05 & 14.36$\pm$0.05 & 14.48$\pm$0.02 & 14.46$\pm$0.04 & 14.44$\pm$0.04 & Caha\\
05/07/01 & 53553.35 & 13.53$\pm$0.06 & 14.36$\pm$0.05 & 14.46$\pm$0.05 & 14.40$\pm$0.02 & 14.34$\pm$0.06 & Ekar\\
05/07/02 & 53554.46 & 13.69$\pm$0.10 & 14.45$\pm$0.04 & 14.51$\pm$0.03 & 14.47$\pm$0.03 & 14.29$\pm$0.10 & Ekar\\
05/07/05 & 53557.42 &                & 14.52$\pm$0.04 & 14.54$\pm$0.04 & 14.42$\pm$0.05 & 14.29$\pm$0.08 & Ekar\\
05/07/06 & 53557.84 & 14.02$\pm$0.06 & 14.60$\pm$0.05 & 14.55$\pm$0.03&  14.36$\pm$0.03 & 14.40$\pm$0.03 & Sub\\
05/07/07 & 53559.40  &                & 14.65$\pm$0.04 & 14.56$\pm$0.09 & 14.41$\pm$0.09 & 14.35$\pm$0.11 & TNT\\
05/07/11 & 53563.38 & 14.81$\pm$0.05 & 14.87$\pm$0.04 & 14.56$\pm$0.04 & 14.37$\pm$0.02 & 14.23$\pm$0.05 & Ekar\\
05/07/11 & 53563.42 & 14.89$\pm$0.04 & 14.93$\pm$0.04 & 14.53$\pm$0.04& 14.37$\pm$0.03 & 14.26$\pm$0.03 & TNG\\
05/07/13 & 53565.38 & 15.27$\pm$0.06 & 15.04$\pm$0.05 & 14.64$\pm$0.02& 14.37$\pm$0.03&  14.29$\pm$0.05&  LT\\
05/07/14 & 53566.36 & 15.29$\pm$0.05 & 15.11$\pm$0.05 & 14.68$\pm$0.02 & 14.44$\pm$0.04 & 14.23$\pm$0.02 & Ekar\\
05/07/14 & 53566.40  &                & 15.09$\pm$0.04 &14.66$\pm$0.04 & 14.40$\pm$0.05 & 14.32$\pm$0.06 & TNT\\
05/07/17 & 53569.42  & 15.92$\pm$0.07&  15.31$\pm$0.05 & 14.67$\pm$0.03 & 14.40$\pm$0.03 & 14.27$\pm$0.04 & LT \\
05/07/19 & 53571.40  &                & 15.36$\pm$0.12 & 14.72$\pm$0.05 & 14.45$\pm$0.09 & 14.26$\pm$0.07 & TNT\\
05/07/20 & 53572.40  &                & 15.39$\pm$0.07 & 14.71$\pm$0.03 & 14.45$\pm$0.03 & 14.27$\pm$0.03 & TNT\\
05/07/25 & 53577.40  &                & 15.46$\pm$0.09 & 14.72$\pm$0.04 & 14.36$\pm$0.06 & 14.24$\pm$0.06 & TNT\\
05/07/27 & 53579.40  &    	      & 15.60$\pm$0.07 & 14.73$\pm$0.04 & 14.39$\pm$0.04 & 14.16$\pm$0.04 & TNT\\
05/07/31 & 53583.39  & 16.83$\pm$0.09 & 15.78$\pm$0.06 & 14.71$\pm$0.03 & 14.38$\pm$0.03 & 14.12$\pm$0.05 & TNG\\
05/07/31 & 53583.47  & 16.81$\pm$0.11 & 15.77$\pm$0.06 & 14.69$\pm$0.03 & 14.36$\pm$0.03 & 14.15$\pm$0.04 & LT\\
\hline
\end{tabular}

NCO = Osservatorio Cavezzo 40cm Newton Telescope\\
Caha = Calar Alto 2.2m Telescope + CAFOS\\
Ekar = Asiago 1.82m Copernico Telescope + AFOSC\\
Sub = Subaru Telescope 8.2m + FOCAS\\
TNG = Telescopio Nazionale Galileo 3.5m + Dolores\\
LT = Liverpool Telescope 2.0m + RATCAM\\
TNT = Teramo Normale Telescope 72cm
%\end{footnotesize}
\end{table*}

\begin{figure}
\resizebox{\hsize}{!}{\includegraphics{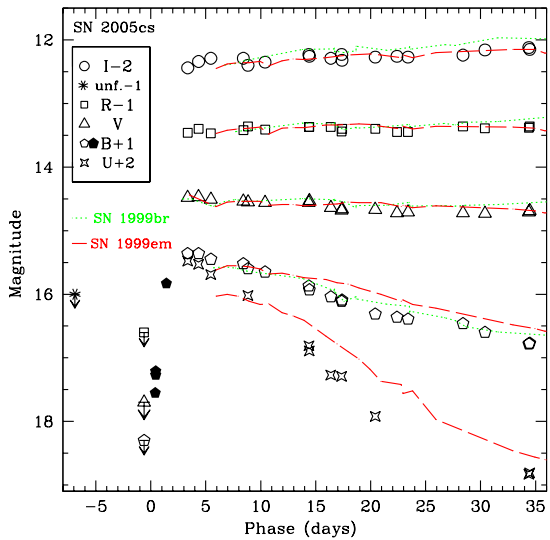}}
\caption{Early U, B, V, R, I light curves of SN 2005cs, including our limits of June
26 (when the SN was not detected) and the unfiltered limit from IAU
Circ. 8853 (asterisk). 
Also the light curves of SN 1999br
(BVRI) and SN 1999em (UBVRI) have been included for comparison, and
shifted in magnitude by an arbitrary amount in order to match the corresponding
curves of SN 2005cs. The very early B band detections of M. Fiedler
(SNWeb) have been also reported (filled pentagons).
}\label{fig:lightcurves}
\end{figure}

The SN magnitudes are reported in Tab. \ref{SN_mags}, and the
U, B, V, R, I light curves are shown in Fig. \ref{fig:lightcurves} together with
the light curves of SNe 1999br \citep{ham01,pasto04} and 1999em
\citep{ham01b,leo02,abou03}, shifted arbitrarily in magnitudes 
(but not in phase) in order to match the light curves of SN 2005cs. 
The overlap of the B, V, R and I band light curves of the three SNe is
very good, while some differences are visible in the U band
evolution: the U band light curve of SN 2005cs declines
more rapidly than that of SN 1999em.\\

According to \citet{modj05}, the simultaneous presence in the SN spectrum
of narrow Galactic and host galaxy interstellar Na ID lines with analogous
equivalent width (0.2 \AA) indicates a similar contribution of the two
components to the total extinction. The host galaxy extinction can be
estimated using the relation of \citet{tura03}, while
the value provided by \cite{schl98} is adopted for the Galactic extinction.
Taking into account both components, a total reddening of E(B--V) = 0.06 
is estimated. However, other methods yield slightly higher values for the
reddening.
The study of the nearby H II region CCM 56 and the three--colour photometry of some
red supergiants close to the SN position yield a total reddening
of  E(B--V) = 0.16 \citep{bres04} and E(B--V) = 0.12 \citep{mau05b},
respectively. Lacking strong constraints in favour of one of these
three methods, an average value of E(B--V) = 0.11$\pm$0.04 is adopted.
However, this relatively low amount of interstellar reddening is consistent 
with the blue colour of the early spectra of SN~2005cs.

\begin{figure*}   
\centering
 \includegraphics[width=15cm]{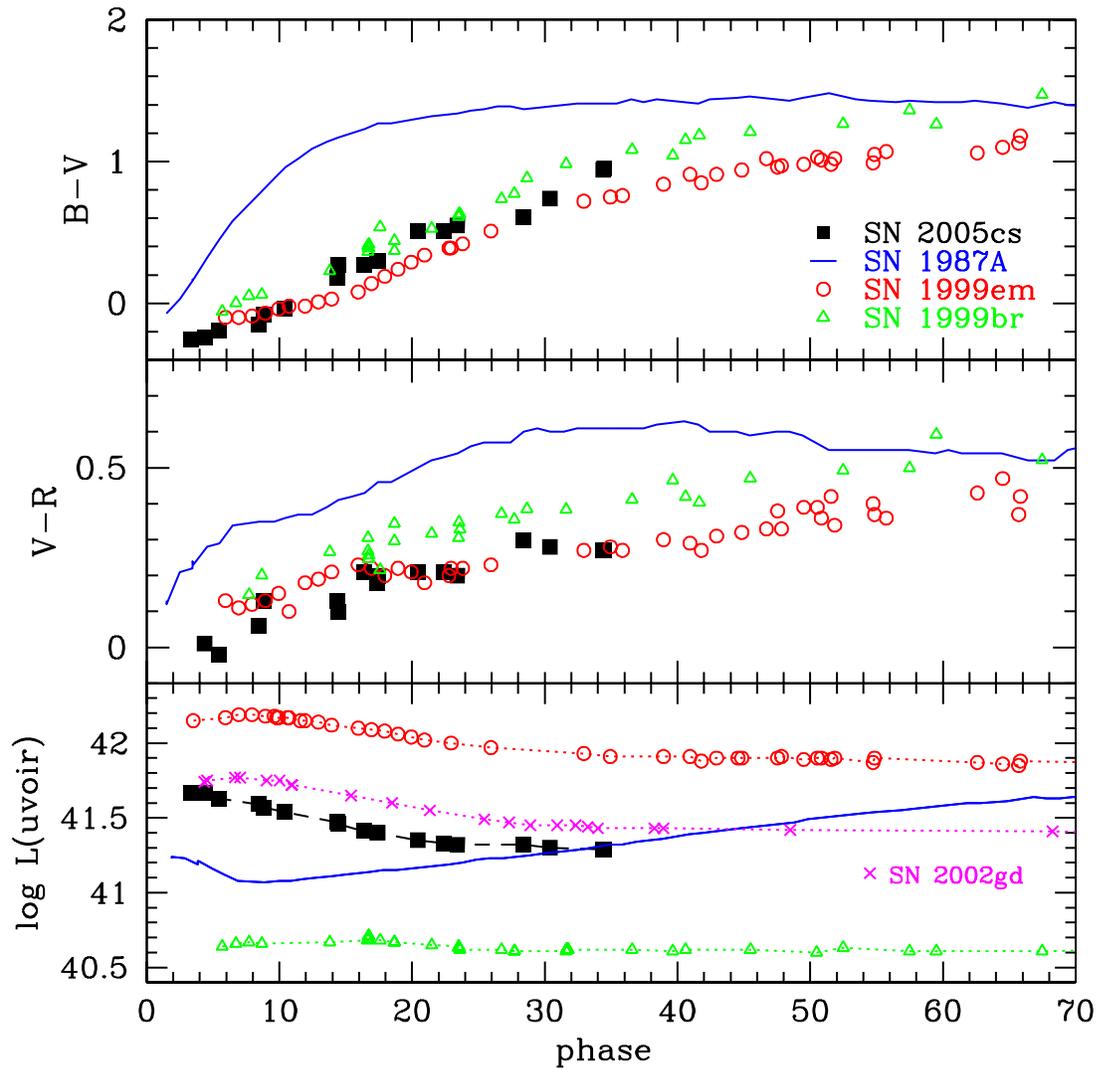}
   \caption{Top: early B--V colour curve of SN 2005cs and those of
SNe 1999em, 1999br 
and 1987A. Middle: same as above, but for V--R colour. Bottom: early 
{\it uvoir} pseudo--bolometric light curves for the same sample of SNe IIP and, in addition, SN~2002gd. 
For references, see the text.}
 \label{fig:color_bolom}
\end{figure*}

\cite{feld97} used planetary nebulae to determine a distance to
M51 of 8.4 Mpc (i.e. distance modulus $\mu$ = 29.62$\pm$0.15). 
Averaging all R band data reported in Tab. \ref{SN_mags}, an R band absolute
magnitude during the plateau phase M$_R$ $\approx-$15.48$\pm$0.16 was
obtained. This implies that SN 2005cs was
relatively underluminous compared to more typical SNe IIP and 
was therefore similar to SNe 1997D and 1999br 
\citep{tura98,bene01,ham01,zamp03,pasto04}, although less extreme.

In Fig. \ref{fig:color_bolom} we compare the early time B--V (top) and V--R (middle) colour
curves, and the pseudo--bolometric {\it uvoir} light curve (bottom) of
SN 2005cs with those of the SNe IIP 1999br \citep{ham01,pasto04}, 
1999em \citep{ham01b,leo02,abou03} and the peculiar 1987A \citep{menz87}. The light curve
of another moderately under--luminous object, SN 2002gd \citep{pasto03}, is also included in
Fig. \ref{fig:color_bolom}(bottom) for comparison.

We adopted a distance modulus $\mu$ = 30.97 and a total
B band extinction A$_B$  = 0.10 \citep{pasto04} for SN 1999br, $\mu$ = 30.34 \citep{leo03}
and  A$_B$  = 0.41 \citep{eddie00} for SN 1999em,  $\mu$ = 18.49 
and  A$_B$  = 0.79 \citep{arn89} for SN 1987A, and
$\mu$ = 32.87 and A$_B$  = 0.29 \citep{pasto03} for SN 2002gd.
Distance moduli for SNe 1999br and 2002gd were estimated from
the recession velocity corrected for the Local Group infall
onto the Virgo Cluster (v$_{Vir}$), adopting 
a value of H$_0$ = 72 km s$^{-1}$ Mpc$^{-1}$.

All typical SNe IIP show a similar colour evolution. 
Their colour curves become monotonically redder with time.
On the other hand, SN 1987A was significantly redder at early 
epochs (until $\sim$ 50 days), but 
then the colour difference between it and other SNe IIP decreases with time.
SN 2005cs follows the behaviour of SNe IIP: the B--V colour increases from
--0.2 to 1 during the first month, while the V--R colour increases
much more slowly (0 to 0.3) over the same period (Fig. \ref{fig:color_bolom} top and middle).

Interestingly, the {\it uvoir} light curve of SN 2005cs has an evolution
reminiscent of that of normal SNe IIP (Fig. \ref{fig:color_bolom}, bottom).
The  {\it uvoir} luminosity of SN 2005cs is intermediate between those of the 
intermediate $^{56}$Ni mass 
SN 1999em  \citep[M$_{Ni}$ $\approx$ 0.05M$_{\odot}$,][]{leo03} and the low--luminosity, $^{56}$Ni--poor 
SN 1999br \citep[M$_{Ni}$ $\approx$ 2$\times$10$^{-3}$M$_{\odot}$,][]{zamp03,pasto04}, 
and is similar to that of the $^{56}$Ni--poor SN 2002gd \citep[see Sect. \ref{discussion} and][]{pasto03}.

\begin{table*}
%\begin{footnotesize}
\caption{Journal of spectroscopic observations of SN 2005cs.
The phase is from the explosion epoch.}\label{spectra}
\centering
\begin{tabular}{cccccc}
\hline\hline
Date & JD & Phase & Instrumental & Range     & Resolution$^{a}$ \\
     & +2400000   & (days) &configuration  & (\AA) & (\AA)  \\ \hline
05/06/30 & 53552.44  & 3 &  CA3.5m+PMAS+grt.V300& 4700--8000  &     9 \\
05/07/01 & 53553.41  & 4 &  Ekar+AFOSC+gm.4,2   & 3500--10100  & 24,38 \\
05/07/02 & 53554.39  & 5 &  Ekar+AFOSC+gm.4,2   & 3500--10100  & 24,38 \\
05/07/02 & 53554.45  & 5 &  Pennar+B$\&$C+150tr/mm & 4100--8700   & 25 \\ 
05/07/05 & 53557.43  & 8 &  Ekar+AFOSC+gm.4,2   & 3650--9900  & 24,38 \\
05/07/06 & 53557.84  & 9 &  Subaru+FOCAS+B300+Y47 & 4800--9000  & 11 \\
05/07/11 & 53563.36  &14 &  Ekar+AFOSC+gm.4   & 3500--7800  &    24 \\
05/07/11 & 53563.44  &14 &  TNG+DOLORES+gm.LRB,LRR & 3150--9500  & 18,17\\
05/07/14 & 53566.35  &17 &  Ekar+AFOSC+gm.4,2   & 3500--9200  & 24,38\\
05/07/15 & 53567.42  &18 &  Pennar+B$\&$C+150tr/mm & 4150--8750   & 25 \\ 
05/07/19 & 53571.45  &22 &  Pennar+B$\&$C+150tr/mm & 4050--7900   & 25 \\ 
05/07/31 & 53583.44  &34 &  TNG+DOLORES+gm.LRB,LRR &3150--9600 & 18,17 \\
\hline
\end{tabular}

$^{a}$ as measured from the full--width at half maximum (\textit{FWHM}) of the night--sky lines\\
CA3.5m = 3.5m Telescope, Calar Alto Obs., Centro Astr. Hispano Alem\'an, Almer\'ia (Spain)\\
Ekar = 1.82m Copernico Telescope, INAF - Osservatorio di Asiago, Mt. Ekar, Asiago (Italy) \\
Pennar = 1.22m Galilei Telescope, Universit\`a di Padova, Loc. Pennar,	Asiago (Italy) \\
Subaru = 8.2m Subaru Telescope, National Astr. Obs. of Japan, Mauna Kea, Hawaii (USA)\\
TNG = 3.5m Telescopio Nazionale Galileo, La Palma, Canary Isl. (Spain)  
%\end{footnotesize}
\end{table*} 

\begin{figure*}   
\centering
 \includegraphics[width=12.5cm]{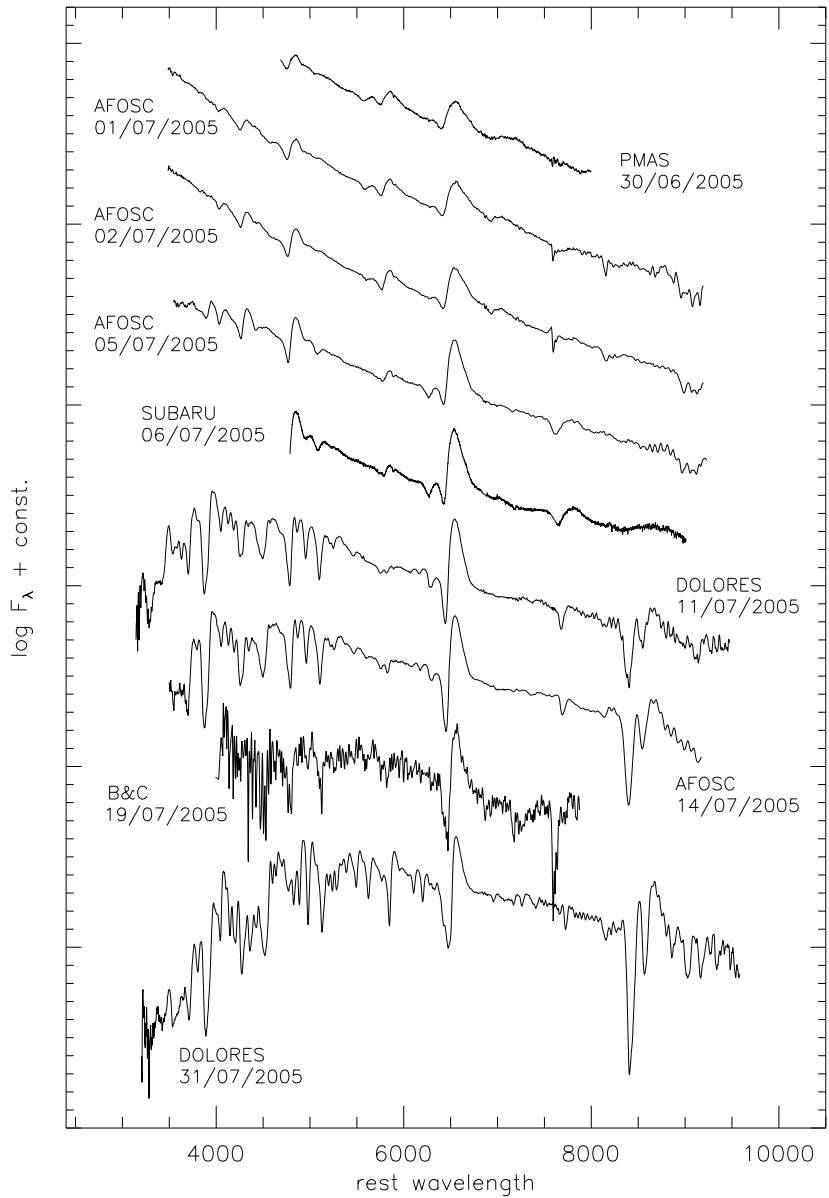}
   \caption{Spectroscopic evolution of SN 2005cs from $\sim$3 days
   to about one month after explosion. Telluric
   features were not removed from the noisy B$\&$C spectrum of
 July 19.  For details, see also Tab. \ref{spectra}. All spectra are 
 at the host galaxy rest frame.}
 \label{fig:spectra}
\end{figure*}

\section{Spectroscopy}

Spectroscopy is available for 10 different epochs, ranging from about 3 to 35
days after the explosion. A summary of all spectroscopic
observations is given in Tab. \ref{spectra}. 

All raw frames were
first bias and flat--field corrected, and then the SN spectra were optimally 
extracted. Wavelength calibration was obtained with the help of
comparison lamp exposures, while the spectra were flux
calibrated using standard star spectra obtained on the same night as
the SN observations. 
When no spectro--photometric standard star was observed, 
a sensitivity function derived on a different night (close in time)
was used.
Telluric features were removed from the SN spectra, again using
spectro--photometric standard spectra. However, the strong
telluric band at 7570--7750\AA\ is superimposed on the O I $\lambda$7774 
SN feature, and imperfect removal might significantly affect the
profile of this line. 
Finally all spectra were checked against the quasi--contemporaneous
photometry and, where discrepancies occurred
(especially during low--transparency or bad--seeing nights), 
the photometric data were used to derive a scaling factor to apply to the SN spectrum.
The relative, final flux calibration was reasonable, 
and the agreement with photometry within 10\%.

\subsection{Spectral evolution}

We monitored SN~2005cs spectroscopically for about one month. 
There are only two significant observational gaps,
between July 6 and 11 (phase 9--14 days) and between July 19 and 31
(phase 22--34). During the last part of the coverage the spectrum
of SN~2005cs evolved significantly.
The spectral sequence is shown in Fig. \ref{fig:spectra}.

The first spectra (phase 3--5 days) are characterized by a very blue continuum.
The most prominent features are the P--Cygni profiles of the H Balmer
lines and of He I 5876\AA. The position of
the minimum of these lines indicates expansion velocities of
the ejecta between 5000 and 8000 km s$^{-1}$, significantly lower than those
typically observed in SNe IIP at a comparable phase \citep[cf. e.g.][]{ham01,ham03,pasto03}.
The most intriguing property is the presence of absorption
features on the blue side of H$\beta$, of He I 5876\AA~ (very
prominent) and, though this is clearly visible only in more evolved
spectra, of H$\alpha$.
These absorptions can be interpreted as either high--velocity (HV) H I and He I features, 
or as absorption lines of other ions (N II and Si II).  
The possible presence of HV lines could be explained either by
an unusual density structure of the progenitor star, with a relatively
dense and He--rich outer shell, or by strong mixing and asphericity
of the ejecta. As discussed in  Sect. \ref{line_id},
there is evidence that these lines are due to N II and Si II.

The subsequent two spectra (phase 8--9 days) also show a blue continuum, but the features
near 4580\AA~ and 5580\AA~ have completely disappeared. The Fe II multiplet 42 lines
($\lambda$4924, $\lambda$5018, $\lambda$5169) begin
to be visible to the red side of H$\beta$. The He I 5876\AA~ line becomes
dimmer and a strong O I $\lambda$7774 line is now visible. Moreover, a very prominent
absorption feature is now well developed at $\sim$6300\AA, close to the blue wing of the
H$\alpha$ absorption. We tentatively identify this line as
the Si II 6347, 6371\AA~ doublet (the feature that is prominent in the photospheric
spectra of SNe Ia, hereafter Si II 6355\AA), rather than as a HV
H$\alpha$ component (see also Sect. \ref{line_id} and Sect. \ref{line_id_sect}).
%This is probably a feature distinct from the high--velocity H$\alpha$ candidate.

The spectra at phases 14--22 days show redder continua and
lines with deeper P--Cygni profiles. 
In the region below $\sim$5300\AA, together with the Balmer H lines,
we identify several metal lines. In addition to strong lines of Fe II, Ti II and Sc II
(e.g. the absorptions at 5250\AA~and 5470\AA),
Sr II $\lambda$4078, $\lambda$4216 (doublet 1) and $\lambda$4162, $\lambda$4305
(doublet 3) are possibly detected. 
Consistently with the lower effective temperature, the He I line is no longer
visible and the feature near 5800\AA\ is probably due to the increasing 
strength of Na ID.
Finally, Ca II H$\&$K and the Ca II IR triplet are now among the most
prominent features.  The components of the Ca II IR triplet are not
blended, confirming the low ejecta velocity.
%Unfortunately the low signal--to--noise ratio of the spectrum at phase
%$\sim$22 days does not allow us to appreciate any significant evolution
%of its features.

By the times of the last TNG observation, obtained about 34 days after the explosion,
the spectrum had noticeably changed. Overall, it
looks like a spectrum of a typical SN II during the recombination phase.
The continuum is much redder, and the flux deficit below $\sim$3800\AA\
is due to strong line blanketing from the Fe II features
\citep{mazz92}, but other metal ions may also contribute significantly 
to the continuum shape. The most prominent lines are now H$\alpha$,
Ca II H$\&$K and the Ca II IR triplet, all with very well developed
P--Cygni profiles.
A detailed line identification of this spectrum is presented in
Sect. \ref{line_id_sect}.

\subsection{High--velocity H I, He I features or N II, Si II lines?} \label{line_id}

\begin{figure} 
 \resizebox{\hsize}{!}{\includegraphics{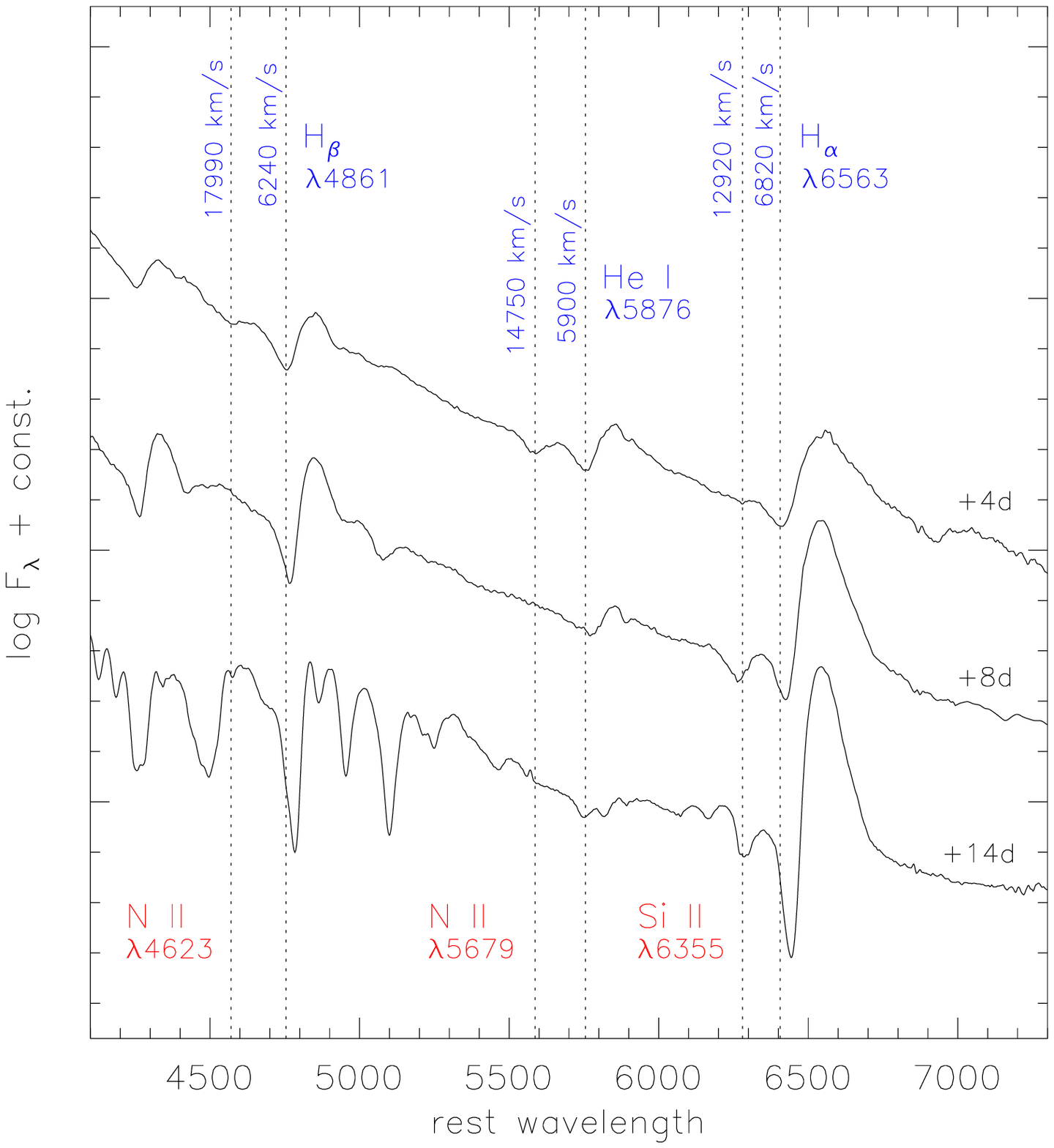}}
   \caption{Evolution of the region between 4100 and 7300\AA,
   where the minima of H$\beta$, He I 5876\AA\ and H$\alpha$
are marked, as well as the position of the features at $\sim$ 4580\AA,
5580\AA\ and 6300\AA.}
   \label{fig:spectra_highv}
\end{figure}

Fig. \ref{fig:spectra_highv} shows the early evolution of the 
spectral region between 4100\AA\ and 7300\AA. 
%evolves at three epochs: $\sim$4 days, $\sim$8 days and $\sim$14 days
%post--explosion. 
The H$\beta$, He I 5876\AA\ and H$\alpha$ absorption minima are marked and the
corresponding line velocities are reported.
The features at 4580\AA, 5580\AA\ and near 6300\AA\ are also marked,
with labelled the two alternative identifications: as putative HV H
and He I (with the corresponding line velocities), and as
N II and Si II lines (bottom of Fig. \ref{fig:spectra_highv}).
\citet{eddie00} discuss the identification of the 4580\AA\ and 5580\AA\ features
in an early time spectrum of SN~1999em. 
Using the simple parametrised code SYNOW \citep{fish00}, 
in which the relative line strengths for each ion are fixed assuming 
local thermodynamic equilibrium (LTE), \citet{eddie00} find that
these features could be consistent with N II $\lambda$4623 and
$\lambda$5679. However, the non-LTE model atmosphere code PHOENIX 
provides a synthetic spectrum which leads them to reject this identification,
because it would require an overabundance of nitrogen to reproduce 
these features. Therefore they support the identification  of the 4580\AA, 5580\AA\ absorptions
as secondary features of H$\beta$ and He I 5876\AA\ at high velocity ($\sim$ 20000
km s$^{-1}$), produced by complicated and unexplained non--LTE effects.
Alternatively, using the model atmosphere code CMFGEN \citep{hill98}
and assuming a relevant N enrichment,   
\citet{luc05,luc06} reproduce the lines in the blue wings
of both of H$\beta$ and He I 5876\AA\ detected in the spectra of SN 1999em as N II.  

\begin{figure}
 \resizebox{\hsize}{!}{\includegraphics{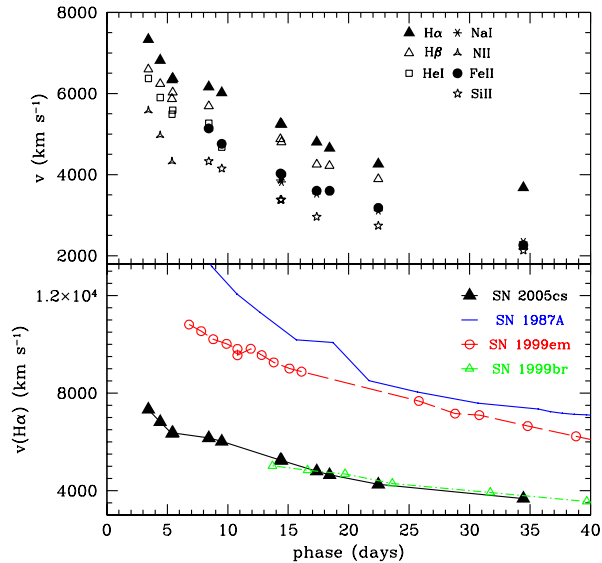}}
   \caption{Top: expansion velocities for H$\alpha$, H$\beta$, 
He I $\lambda$5876, Na ID, N II $\lambda$5680, Fe II $\lambda$5169, Si II $\lambda$6355 
deduced from the minima of P--Cygni profiles in the spectra of
   SN~2005cs.
Bottom: comparison of the H$\alpha$ velocity evolution for SN~2005cs,
   SN~1987A, SN~1999em and SN~1999br. See the text for references.}
   \label{fig:velocities}
\end{figure}

For SN~2005cs a scenario where two line forming regions exist
for hydrogen and helium is not supported by the line 
velocities measured for the putative HV components. As shown in
Fig. \ref{fig:spectra_highv}, these line velocities are
inconsistent.
In particular, the velocity of the putative HV H$\beta$ component
is much larger than that of the HV H$\alpha$ component.
Moreover, the putative HV H$\beta$ and HV He I
features disappear simultaneously in the spectrum at $\sim$8 days.
This supports, at least in the case of SN 2005cs, 
the identification of these two features as both due to N II. 

For the same reason, and because of its persistence over a much 
longer time than the feature at 4580\AA, we believe that the feature 
near 6300\AA\ is Si II 6355\AA, rather than HV H$\alpha$. 
The presence of this absorption in the spectra of SN~1999em
was claimed by \citet{luc05}, and not explicitly mentioned by \citet{eddie00}.

In Fig. \ref{fig:velocities}(top) we show the evolution of the 
velocity of various absorption spectral lines in SN~2005cs, as derived from the position
of their minima. H$\alpha$ shows the highest velocities, about 1000 km
s$^{-1}$ larger than those of the He I line. The N II
and the Si II features appear to have smaller velocities than other
lines. In particular, the velocity 
of the N II lines decreases from about 5600 km s$^{-1}$ to $\sim$3200
 km s$^{-1}$ between $\sim$3 and 5 days. 
%systematically about 1000  km s$^{-1}$ below the He I 5876\AA\
%line and 2000 km s$^{-1}$ below H$\alpha$.
The velocity of the Si II 6355\AA\ feature 
decreases from 4300 km s$^{-1}$ at $\sim$8 days to about 2100 km s$^{-1}$ one
month after the explosion. Over the same time interval, the H$\alpha$ velocity decreases
from 6200 to 3700 km s$^{-1}$, while the Fe II velocity is
$\sim$800 km s$^{-1}$ faster than that of the Si II line. 
In  Fig. \ref{fig:velocities}(bottom) we compare the
 H$\alpha$ velocity evolution in SN~2005cs with that of SN 1987A
\citep{phil88}, SN~1999em \citep{pasto03} and SN~1999br
\citep{ham01,pasto04}. The H$\alpha$ velocity curve of SN~2005cs
appears to be strikingly similar to that of the low--velocity SN~1999br.

\subsection{Spectral models: the 17 and 34 days spectra} \label{line_id_sect}

Some preliminary spectral models have been computed in order to provide
basic line identification. The code used for the synthetic spectra 
was described in more detail by
\citet{abbott85,mazzali93a,lucy99,mazzali00}.  The procedure involves a
Monte Carlo (MC) simulation of the line transfer based on the Sobolev
approximation. The code assumes that all radiative energy is emitted
below a sharp lower boundary.  The propagation of all energy packets is followed
through the spherically symmetric envelope. Processes of interaction
for photons taken into account are electron scattering and line
transitions. When a photon packet is absorbed by a line transition, it is
reemitted at a new frequency corresponding to the branching probabilities
for the radiative decays of the excited level. At the end of the MC
calculation a formal integral routine derives the emergent spectrum based
on the MC estimate of the source functions \citep[see][]{lucy99}.

\begin{table}
  \caption{Model parameters for synthetic spectra shown in Fig. \protect\ref{fig:line_id_daniel}}
  \label{table:model_paramter}
  \centering
  \begin{tabular}{lcc}\hline\hline
                        & $t=17^{d}$ & $t=34^{d}$  \\ \hline
    $\log(L/L_{\odot})$         &  7.90   & 7.88            \\
    $v_{\rm ph}$ [km\,s$^{-1}$] &  3710   & 2580            \\
    $\log(R_{\rm ph}/R_{\odot})$ &  3.894  & 4.037            \\
    $T_{\rm ph}$ [K]            &  7599   & 6223            \\ \hline
  \end{tabular}
\end{table}

\begin{figure*} 
\centering
 \includegraphics{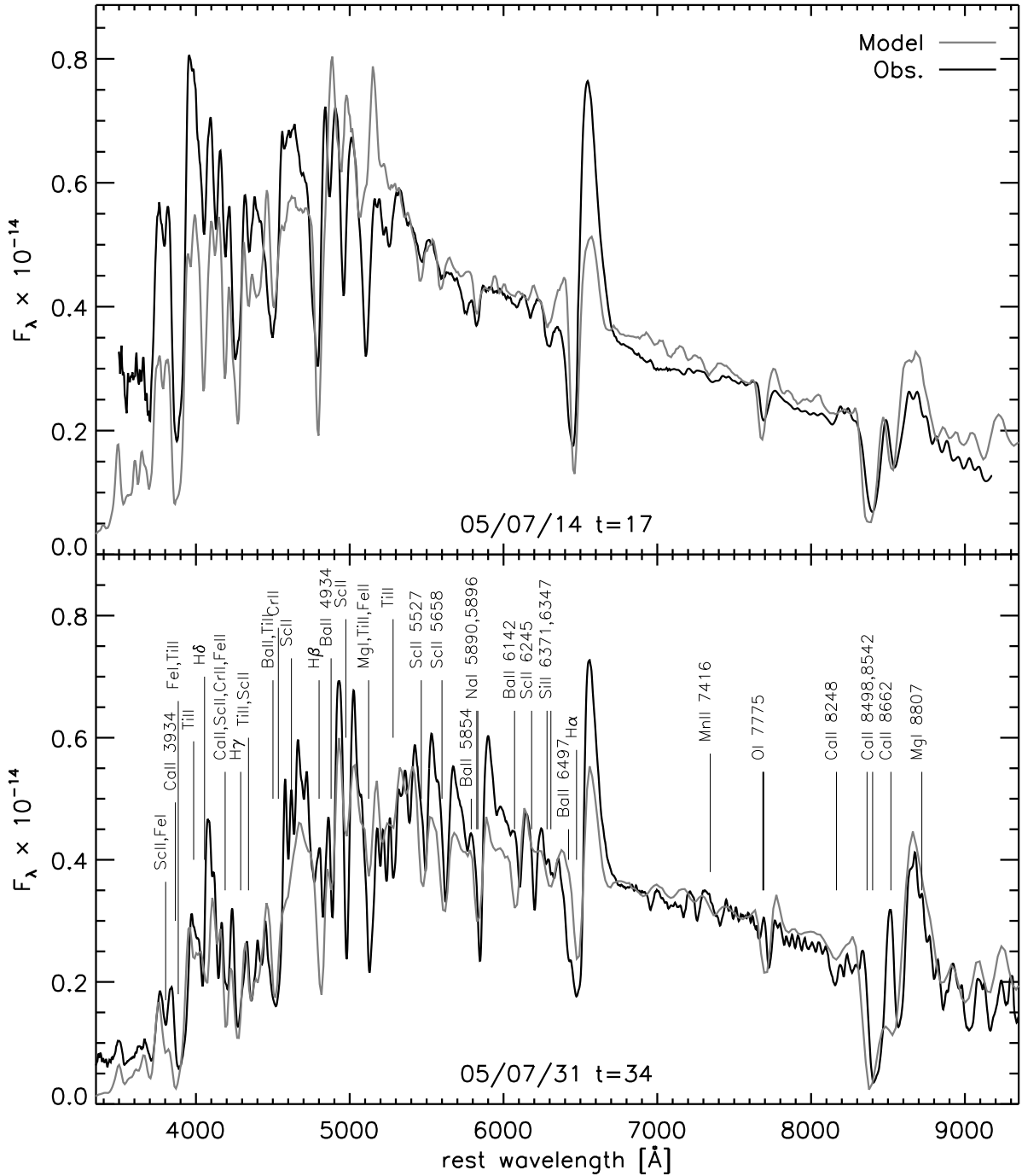}
   \caption{Comparison between synthetic spectra and the
   observed spectra of SN~2005cs at phase 17 (top) and 34 days
   (bottom). The parameters used to compute the synthetic spectra are
   reported in Tab. \protect\ref{table:model_paramter}.}
   \label{fig:line_id_daniel}
\end{figure*}

%The global parameters adopted for the spectral models are the distance
%modulus to M51 of ${m-M=29.6}$ \citep{feldmeier97}, the suggested total
%reddening for the supernova  $E(B-V)=0.08$; the explosion date is assumed
%to be June 28th 2005. 
All models were computed assuming solar
composition \citep{gre98}.  As a starting point for this preliminary analysis we used
the density structure of the hydrodynamic explosion model
adopted to fit SN 1997D \citep{tura98}.  
This model was characterised by an outer density profile which can be
approximated by a power law $\rho \propto r^{-10}$ at velocities above 3000
km s$^{-1}$, and contained less than $\sim$ 0.1$M_{\odot}$ of material above this
velocity.
However, in order to reproduce
the relatively narrow absorptions, in particular the well separated lines
of the Ca II IR triplet, the density structure was cut above
$v=6500\,$km\,s$^{-1}$ by adopting an even steeper power law 
($\rho\sim r^{-40}$) beyond that velocity.
This modification leads to an insignificant reduction of the ejected mass. 
Tab. \ref{table:model_paramter} gives an overview
on the model parameters used to derive the synthetic spectra shown in
Fig. \ref{fig:line_id_daniel}. 
Input parameters include the total luminosity of the
SN and the position of the
photosphere in velocity space $v_{\rm ph}$. Together with the epoch $t$
this constrains the photospheric radius $R$ and, therefore, the
temperature at the photosphere via $T_{\rm ph}^{4}=L/4\pi\,\sigma
R^{2}=L/4\pi\,\sigma t^{2}v_{\rm ph}^{2}$, where $\sigma$ is the
Stefan--Boltzmann constant.  Tab.~\ref{table:model_paramter} gives the
final temperature at the photosphere which is determined iteratively to
obtain the required emergent luminosity at the outer radius. 

The synthetic spectra in Fig. \ref{fig:line_id_daniel} are
compared with the observed spectra at phases $\sim$17 (top) and 34
days (bottom), respectively.
The bottom panel of Fig. \ref{fig:line_id_daniel} also shows the identifications of
prominent line features in this spectrum. The synthetic spectra support
the identification of the feature near $6300\,${\AA} with the Si II
6347, 6371\AA\ lines. 
In the region between 5000 and 6400\AA, together with strong Fe II features,
prominent lines of Sc II, Ti II, and the Na ID lines are identified.

In the later spectrum we note the increased strength of the Ba II lines
compared to earlier epochs. 
 Unfortunately, the 5854\AA\ line is blended with Na ID and
the 6497\AA\ line with H$\alpha$. However, the feature at 6142\AA\ is
relatively unblended and is detected unambiguously.
%In particular the H$\alpha$ line is broadened by the influence of
%the \ion{Ba}{ii} $\lambda\,6497$ line.
The Ba II lines, as well as those of other s--process elements (Sc, Sr), 
are particularly prominent in 
late--photospheric spectra of low--velocity SNe IIP \citep{pasto04}
and of SN 1987A, while are weaker in other SNe IIP \citep{mazz92,mazz95,utro05}.

The proposed  density structure does not seem to be able to fit
the observations in detail. In particular, it is apparent that the
proposed density cut, which is required to fit the Ca II IR lines at
$t=17\,$d, does not provide a good fit for the epoch $t=34\,$d.  This
suggests that a steeper density law with less mass at high
velocity is required to reproduce the low--velocity narrow absorptions
seen in the observed spectra.
%%Attempts to model the very early spectra where not successful
%%because the adopted method does not treat the continuum and the
%%temperature structure consistently. The approximations which have been
%%used, however, are not suitable to describe this very early phase. The
%%shape of the continuum at the first few spectra suggest fairly large
%%temperatures. On the other hand, the apparent presence of \ion{He}{i}
%%lines in the spectrum indicates a region of low degree of ionization which
%%may suggest the presence of non-LTE effects which cannot be treated with
%%the codes employed for these models.  Also the the low velocities of the
%%lines absorptions hints toward significantly lower mass at higher
%%velocities than present in the proposed density structure of SN 1997D.

Concerning the total ejected mass,  no conclusion can be drawn on the basis
of spectral models of early epochs alone.  A more detailed analysis of the
density structure would also require information on the later spectral
evolution as well as a study of the light curve, including the
duration of the plateau phase \citep[e.g.][]{tura98}. This will be
discussed elsewhere.

\subsection{Comparison with other SN IIP}
 
\begin{figure}
 \resizebox{\hsize}{!}{\includegraphics{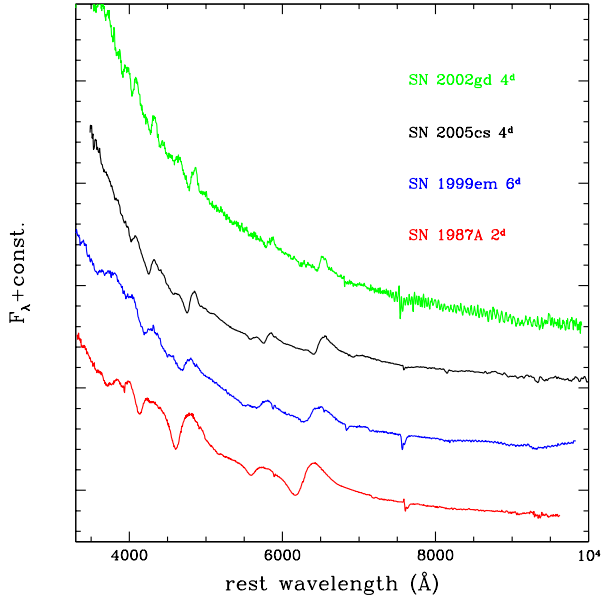}}
   \caption{Comparison among spectra of SN 2002gd, SN 2005cs, SN
   1999em and SN 1987A shortly after explosion.
   For references, see the text.}
   \label{fig:spectra_comp_expl}
\end{figure}

\begin{figure}
 \resizebox{\hsize}{!}{\includegraphics{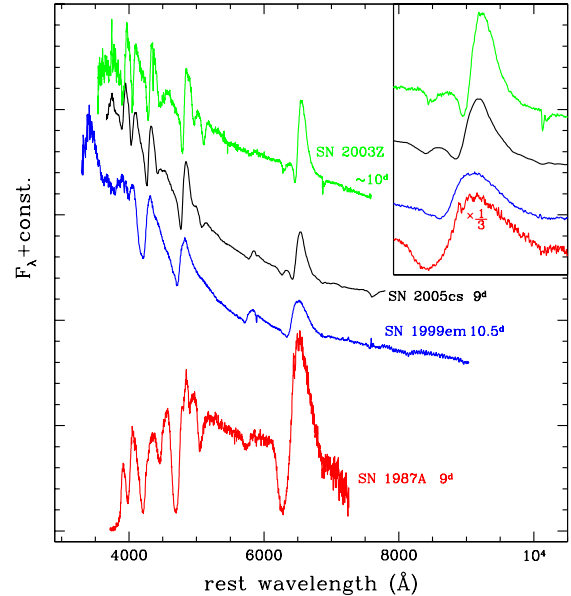}}
   \caption{Comparison among spectra of SN 2003Z, SN 2005cs, SN 1999em and SN
   1987A at a phase of $\sim$10 days. A blow--up of the H$\alpha$ region is shown in the top--right corner, showing the different strengths
 of the Si II feature in SN IIP spectra.}
   \label{fig:spectra_com_10d}
\end{figure}

\begin{figure}
 \resizebox{\hsize}{!}{\includegraphics{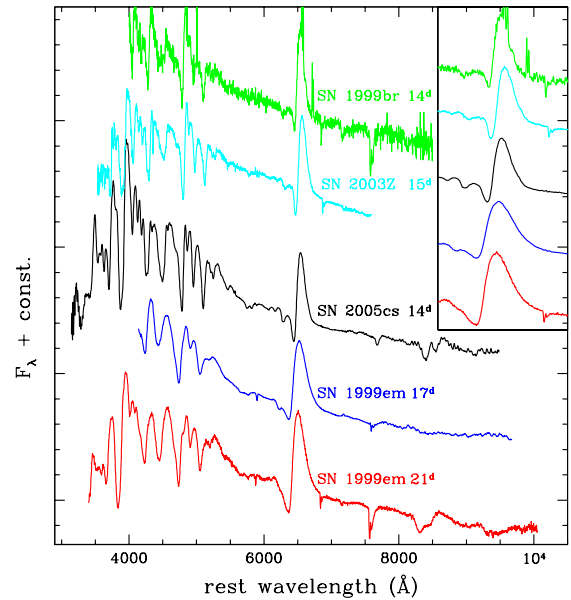}}
   \caption{Spectra of the low--velocity SN IIP 1999br, 2003Z and 2005cs at a phase of about two
   weeks. Spectra of SN
   1999em at a phase of 17 and 21 days are included for
   comparison. Again, a blow--up of the H$\alpha$ region is also shown in the top--right corner.}
   \label{fig:spectra_comp_14d}
\end{figure}

In Fig. \ref{fig:spectra_comp_expl} we compare the spectrum of SN~2005cs at $\sim$4 days
with those of other young SNe IIP: the low--velocity, $^{56}$Ni--poor SN 2002gd \citep{pasto03}, 
SN 1999em \citep{ham01b} and SN 1987A (Padova--Asiago SN archive).
All spectra have blue continua and show He I 5876\AA~ and the H Balmer
lines.
Moreover, the features we attribute to N II 4623\AA~and 5679\AA~ seem
common to all SN II spectra except SN~1987A, which has however
broader lines.
  
At $\sim$10 days after explosion (see Fig. \ref{fig:spectra_com_10d}),
we note a number of significant differences between the spectra of SNe IIP.
While the spectrum of SN~1987A shows a red continuum dominated by
prominent H and Fe II P--Cygni features, the spectra of more typical
SNe IIP are still relatively blue. Although the spectrum of SN~2005cs 
is slightly redder than that of SN~1999em \citep{abou03}, it is bluer 
than that of the low--velocity, $^{56}$Ni--poor SN 2003Z \citep{kno05}.
The N II features visible in Fig. \ref{fig:spectra_comp_expl} 
have now completely disappeared, but the prominent absorption feature
at 6300\AA~ is now visible in the spectra of SNe 2003Z, 2005cs and,
as a bump in the H$\alpha$ absorption profile, in 1999em \citep[see
discussion in][]{abou03}.
The identification of this feature as Si II 6355\AA~is supported by
the spectral modelling presented in Sect. \ref{line_id_sect}.

In Fig. \ref{fig:spectra_comp_14d} we compare the spectrum of SN
2005cs at phase $\sim$ 2 weeks to those of the $^{56}$Ni--poor SNe
1999br \citep{ham01,pasto04} and 2003Z \citep{kno05} at a similar
epoch. We also include 2 spectra of SN 1999em \citep{leo02,ham01b}, taken about 17 and
21 days after the explosion. 
The Si II  6355\AA~ feature is prominent in all spectra,
included that of SN 1999em at phase $\sim$17 days, but it is not longer visible 
4 days later.
  
\section{Discussion} \label{discussion}

\begin{table*}
\caption{Main data about low--luminosity, $^{56}$Ni--poor SNe IIP
available in literature. In column 5 we reoprt the expansion velocity of the
ejecta at phase 
$\sim$35 days, obtained
from the position of the H$\alpha$ P--Cygni absorption.}\label{LL-SNIIP}
\centering
\begin{tabular}{ccccccc}
\hline\hline
SN name & Host Galaxy & Galaxy Type$^{\circledast}$ & M$_{V,pl}$$^\star$ & v$_{35}^{H\alpha}$ (km
s$^{-1}$) & $^{56}$Ni Mass (M$_\odot$)$^\star$ & Source \\
\hline
1994N & UGC 5695  & Sab        & -14.8        & 4250 & 5$\times$10$^{-3}$ & $\blacklozenge$ \\
1997D & NGC 1536  & SBc        & -14.3$^\dag$ & --   & 7$\times$10$^{-3}$ & $\blacklozenge$ \\
1999br& NGC 4900  & SBc        & -13.5        & 3700 & 2$\times$10$^{-3}$ & $\blacklozenge$ \\
1999eu& NGC 1097  & sBb        & -13.7$^\dag$ & --   & $\leq$3$\times$10$^{-3}$ & $\blacklozenge$ \\
2000em& LEDA 143614& ?         & -16.3$^\ddag$& -- & unknown & $\bigstar$ \\
2001dc& NGC 5777  & Sb         & -14.1        & --   & 5$\times$10$^{-3}$ & $\blacklozenge$ \\
2002gd& NGC 7537  & Sbc        & -15.6        & 4100 & $\leq$3$\times$10$^{-3}$ & $\blacksquare$ \\
2003Z & NGC 2742  & Sc         & -14.6        & 4000 & 5$\times$10$^{-3}$ & $\blacksquare$ \\
2004cm& NGC 5486  & Sm         & -13.9$^\ddag$& -- &  unknown & $\ast$ \\
2004eg& UGC 3053  & Sc         & -14.7:$^\ddag$& --  & unknown & $\blacktriangle$ \\
2005cs& M 51      & Sbc        & -15.2        & 3700 & unknown & $\blacktriangledown$ \\
\hline
\end{tabular}

$\blacklozenge$ \protect\cite{pasto04};
$\bigstar$ \protect\cite{stro00};
$\blacksquare$ \protect\cite{pasto03};
%$\ast$ \protect\cite{conn04};
$\blacktriangle$ \protect\cite{youn04,fili04};
$\blacktriangledown$ this paper. \\
$\ast$ {\it http://cheops1.uchicago.edu/pub/snefchart/sneCand-53088-1324-491-run003712-20-5-0189-00031.html};
see also \protect\cite{conn04}.\\
$^\circledast$ LEDA;
$^\dag$ Absolute magnitude measured at the end of plateau;
$^\ddag$ Unknown phase (but during plateau); R band absolute magnitudes.

$^\star$ $\mu$ computed assuming H$_0$ = 72 km s$^{-1}$ Mpc$^{-1}$;
when possible, M$_{V,pl}$ was computed  about 1 month after the explosion.

\end{table*} 

The observational analogies between SN 2005cs and objects similar to SN 1997D \citep{tura98}
are remarkable. In Tab. \ref{LL-SNIIP} we report some significant information
available in the literature for a number of objects belonging to this group.
Only SNe with very low ejecta velocities and/or small ejected $^{56}$Ni
mass ($\le$ 10$^{-2}$ M$_\odot$) are included in Tab. \ref{LL-SNIIP}.
All these objects belong to the faint tail of the luminosity distribution of SNe IIP
\citep{ham01,pasto03,ham03,pasto05a,zamp05} and have similarly low expansion velocities. 
In particular, the good match between the H$\alpha$ velocity curves of SNe~2005cs
and 1999br (Fig. \ref{fig:velocities}, bottom),
provides further support to our idea that SN~2005cs can be regarded as
another SN 1997D--like event. In analogy to other low--luminosity 
SNe IIP \citep{pasto05}, we therefore expect that SN 2005cs will also evolve through
a long plateau (3--4 months) and reach a faint late--time
luminosity. This would be indicative that a very small mass of $^{56}$Ni (of the order of
$\sim$ 10$^{-2}$M$_{\odot}$ or less) was ejected in the explosion,
as the other similar SNe of Tab. \ref{LL-SNIIP}.
%Most of them exploded in late--Type galaxies (mainly of morphological Type Sb--Sc, by
%LEDA\footnote{\it http://leda.univ-lyon1.fr/}), which would be 
%consistent with massive progenitors.

\begin{figure} 
 \resizebox{\hsize}{!}{\includegraphics{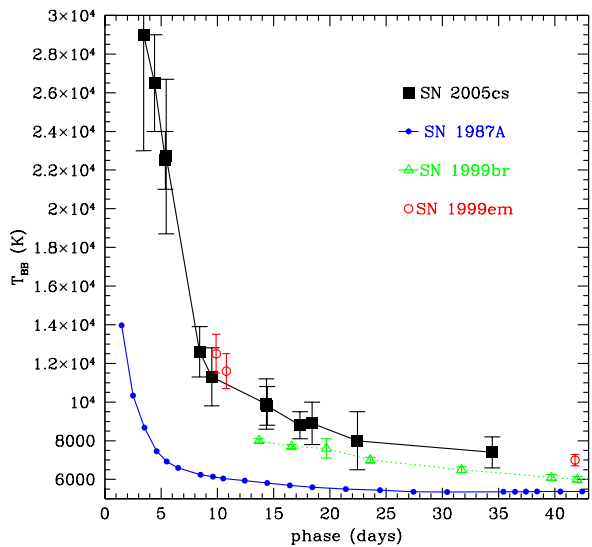}}
   \caption{Evolution of the continuum temperature of SNe 2005cs,
   1987A, 1999em, 1999br.
For references, see text.}
   \label{fig:spectra_temperature}
\end{figure}

The earliest spectra of SN~2005cs are very blue, suggesting a very high continuum
temperature (2--3$\times$10$^4$ K).
In Fig. \ref{fig:spectra_temperature} the evolution of the temperature
of SN~2005cs derived from a blackbody fit to the spectral continuum is compared with those of
SN~1987A \citep{phil88},
SN~1999em \citep[as measured in][]{pasto03} and SN~1999br
\citep{pasto04}.
Again, the continuum temperature evolution of SN~2005cs is not different 
from that of other SNe IIP. In particular, SN~2005cs has
higher continuum temperatures than SN~1999br, but slightly lower than
SN~1999em. Independent on the phase, all these long--plateau
SNe have significantly higher continuum temperatures than SN~1987A.
The temperature of SN~1987A becomes stationary about 3 weeks after
the explosion, while for other SNe IIP the same happens 30--40 days past explosion.
This indicates that in SN~2005cs (and in the other SNe
IIP) the H envelope starts to recombine later than in SN~1987A.  

If the faint absolute luminosity and the very well
developed plateau should be confirmed by the
observational campaign still in progress, they would fit with some difficulty
the relatively small mass (M$_{ZAMS} \sim$ 9M$_\odot$) progenitor scenario
derived from the progenitor detection \citep{mau05b,li05b}. As
mentioned in Sect. \ref{intro}, both groups argued that the low luminosity of 
the progenitor indicates that it had a small mass. \citet{mau05b}
obtained a likely range of bolometric 
luminosity 4 $\la$ log(L/L$_\odot$) $\la$ 4.4 
(shaded region of their Fig. 3), and even for conservative errors, 
log(L/L$_\odot$) $\la$ 4.6. 
\citet{li05b} obtained an even lower value. To
compare with the observed luminosity, they used the Geneva
evolutionary models.  The highest luminosities attained in the Geneva
models are log(L/L$_\odot$) = 4.2, 4.5, and
4.8 for M = 7M$_\odot$, 9M$_\odot$, and 12M$_\odot$, respectively.  
To be consistent with the observed luminosity for the conservative
limit, they suggested M = 7--12M$_\odot$ for the progenitor.

However, the Geneva evolutionary models do not reach the pre--supernova
stage. They cover the evolution up to the formation of 
O+Ne+Mg core for M $\ge$ 9M$_\odot$ and only up to the formation of the C+O core
for M = 7M$_\odot$.  Stars with M $<$ 8M$_\odot$ form a
degenerate C+O core whose mass M$_{core}$ increases towards the
Chandrasekhar mass, M$_{ch}$, as the star climbs the AGB \citep{pacz70}.  Eventually
the stars will either lose their H--rich envelope to form C+O white dwarfs
or undergo a thermonuclear explosion (the so--called Type I--1/2 supernovae)
when M$_{core}$ gets close to M$_{ch}$ \citep[for
reviews, see][]{sugi80,ken88}. The 8--10M$_\odot$ stars form a degenerate
O+Ne+Mg core whose M$_{core}$ also increases toward M$_{ch}$ on the AGB
\citep{ken84}.  Their final fate is also either an O+Ne+Mg white
dwarf or a core-collapse supernova when M$_{core}$ gets close to M$_{ch}$
\citep{ken84,ken87}.

The luminosity L of AGB stars with M $\la$ 10M$_\odot$ obeys
Paczynski's (1970)~ M$_{core}$--L relation.  
If a star with M $\la$ 10M$_\odot$ reaches M$_{core}$ = 1.4M$_\odot$, the
pre--supernova luminosities should be at least 
log(L/L$_\odot$) = 4.8 for X(H) = 0.7, and even higher if the He abundance is
higher \citep[e.g. because of mixing,][]{has93}. If the progenitor was as massive
as 12M$_\odot$, the pre--supernova luminosities would be 
log(L/L$_\odot$) $\ga$ 4.8. Systematic studies of the progenitor luminosity,
including the effect of rotation, 
would be desirable to obtain a more reliable detailed comparison. 

Therefore, the observed luminosity of the putative progenitor of
SN~2005cs is inconsistent
with the pre--supernova luminosity of 7--12M$_\odot$ stars.  If 
large extinction causes the observed luminosity of the progenitor to be very small, 
the observed luminosity cannot be used to constrain the progenitor's mass.
One possibility is that the absolute magnitude (and hence the mass) of the progenitor star
is underestimated because of dust enshrouding the supergiant progenitor and
later swept away by the SN explosion \citep{grah86}.
This scenario was ruled out by \citet{mau05b}
because no K band excess in the magnitude of the progenitor was detected in archival images.
However, we cannot exclude the possibility that a particular dust composition and
an extremely low temperature cause the light to be absorbed at 
optical wavelengths and re--emitted mostly in the mid-- and far--IR
bands \citep[see e.g.][]{pozzo04}.
Dust was observed in the late evolution of another well studied SN IIP,
2003gd \citep{maggie04}. Despite having a plateau luminosity and an 
expansion velocity evolution similar to
the ``normal'' SN 1999em, SN 2003gd showed low late time luminosity,
indicating the ejection of a small mass of $^{56}$Ni (0.015 M$_\odot$),
only a factor two more than SN 1997D. \citet{van03} and
\citet{smar04} found
a moderate--mass progenitor for SN 2003gd (8 M$_\odot$). However, an evident light echo was
recently detected in HST images of this object \citep{van05,sug05}. This was due to SN light
scattered by large--grain dust, located 110--180 pc in front of the
SN, that
survived the initial UV--flash. If this material was present before the SN
explosion, it could be responsible for significant extinction of the star
light, leading to an underestimate of the luminosity (and hence of the mass)
of the progenitor. 

As further support to this discussion, most SNe reported in Tab. \ref{LL-SNIIP} exploded in
late Type galaxies (mainly of morphological Type Sb--Sc, by
LEDA\footnote{\it http://leda.univ-lyon1.fr/}), which would be 
consistent with massive progenitors.

Finally, there is some ambiguity as regards the location of the putative
progenitor of SN~2005cs. 
%as indicated by the alternative progenitor detection of \citet{ric05}.
Therefore, even if the works of \citet{mau05b} and \citet{li05b} seem to
support a moderate mass scenario for the precursor star of SN~2005cs,
late time observations of the explosion site will probably be required in order
to remove the residual uncertainty in the correct identification
of the progenitor.

\section*{Acknowledgements}
This work has been supported by the Italian Ministry for
Education, University and Research (MIUR) under PRIN 2004029938.
The observational campaign has been coordinated by the Italian Intensive
Supernova Program (IISP) ({\it http://web.pd.astro.it/supern/iisp/}).\\
This paper is based on observations collected at the
Centro Astron\'omico Hispano Alem\'an (Calar Alto, Spain),
Asiago and Collurania INAF Observatories (Italy), 1.22m Galilei Telescope
of the Universit\`a di Padova (Asiago, Italy), Subaru Telescope
(National Astronomical Obs. of Japan, Mauna Kea, USA),
Telescopio Nazionale Galileo and Liverpool Telescope (La Palma, Spain).\\
AP is grateful to M. Turatto, W. Hillebrandt and S. J. Smartt for useful discussions.
We also thanks the team of the Osservatorio Astronomico di Cavezzo (Modena, Italy)
and M. Barbieri (INAF -- Osservatorio Astronomico di Padova) 
for providing pre--discovery images of the SN 2005cs site.
We thank the resident astronomers of Telescopio Nazionale Galileo (in
particular Vania Lorenzi), the
Liverpool Telescope, and the 2.2m and 3.5m telescopes in Calar Alto for performing
the follow--up observations of SN~2005cs. We also thank 
Elena Mazzotta Epifani, Roberto Nesci and Giovanna Temporin for the ToO observations
of 2005cs at the 1.82m telescope of Asiago.
We are grateful to E. Baron for providing spectra of SN 2003Z before
publication and to the SNWeb observers for useful information about 
very early observations of SN~2005cs.\\
This research has made use of the NASA/IPAC Extragalactic
Database (NED) which is operated by the Jet Propulsion Laboratory,
California Institute of Technology, under contract with the National
Aeronautics and Space Administration. We also made use of the Lyon-Meudon
Extragalactic Database (LEDA), supplied by the LEDA team at
the Centre de Recherche Astronomique de Lyon, Observatoire de Lyon.\\

\bibliographystyle{mn2e}
\bibliography{biblio}
\end{document}